\documentclass[longbibliography,twocolumn,amsmath,amssymb,amsfonts,superscriptaddress,floatfix,showpacs]{revtex4-1}
\usepackage[utf8]{inputenc}
\usepackage[english]{babel}
\usepackage[colorlinks=true,citecolor=green,linkcolor=red,urlcolor=blue]{hyperref}
\usepackage{braket}
\usepackage{bbm}
\usepackage{graphicx}
\usepackage{color}

\newcommand{\Fig}[1]{Fig.~\ref{#1}}

\newcommand{\skippart}[1]{}
\newcommand{\simplesection}[1]{\emph{#1} ---}

\begin{document}

\title{Mapping topological to conformal field theories through strange correlators}
\author{Matthias~\surname{Bal}}
\affiliation{Department of Physics and Astronomy, Ghent University, Krijgslaan 281, S9, B-9000 Ghent, Belgium}
\author{Dominic~J.~\surname{Williamson}}
\affiliation{Vienna Center for Quantum Technology, University of Vienna, Boltzmanngasse 5, 1090 Vienna, Austria}
\affiliation{Department of Physics, Yale University, New Haven, CT 06520-8120, USA}
\author{Robijn~\surname{Vanhove}}
\affiliation{Department of Physics and Astronomy, Ghent University, Krijgslaan 281, S9, B-9000 Ghent, Belgium}
\author{Nick~\surname{Bultinck}}
\affiliation{Department of Physics, Princeton University, Princeton, NJ 08540, USA}
\author{Jutho~\surname{Haegeman}}
\affiliation{Department of Physics and Astronomy, Ghent University, Krijgslaan 281, S9, B-9000 Ghent, Belgium}
\author{Frank~\surname{Verstraete}}
\affiliation{Department of Physics and Astronomy, Ghent University, Krijgslaan 281, S9, B-9000 Ghent, Belgium}
\affiliation{Vienna Center for Quantum Technology, University of Vienna, Boltzmanngasse 5, 1090 Vienna, Austria}

\begin{abstract}
We extend the concept of strange correlators, defined for symmetry-protected phases in [\href{http://dx.doi.org/10.1103/PhysRevLett.112.247202}{You et al., Phys. Rev. Lett. 112, 247202 (2014)}], to topological phases of matter by taking the inner product between string-net ground states and product states. The resulting two-dimensional partition functions are shown to be either critical or symmetry broken, as the corresponding transfer matrices inherit all matrix product operator symmetries of the string-net states. For the case of critical systems, those non-local matrix product operator symmetries are the lattice remnants of topological conformal defects in the field theory description. Following [\href{http://stacks.iop.org/1751-8121/49/i=35/a=354001}{Aasen et al., J. Phys. A 49, 354001 (2016)}], we argue that the different conformal boundary conditions can be obtained by applying the strange correlator concept to the different topological sectors of the string-net obtained from Ocneanu's tube algebra. This is demonstrated by calculating the conformal field theory spectra on the lattice in the different topological sectors for the Fibonacci (hard-hexagon) and Ising string-net. Additionally, we provide a complementary perspective on symmetry-preserving real-space renormalization by showing how known tensor network renormalization methods can be understood as the approximate truncation of an exactly coarse-grained strange correlator.
\end{abstract}
\maketitle

\simplesection{Introduction} 
The mathematical formalism of tensor fusion categories~\cite{etingof2015tensor} appears naturally in both conformal field theory (CFT)~\cite{Moore1989,segal1988definition} and topological quantum field theory (TQFT)~\cite{Witten1988,atiyah1988topological}. Tensor fusion categories describe ``symmetries" in the system under consideration which cannot be captured by group theory as they are not necessarily tensor products of local actions. The connection between TQFTs and CFTs generally arises from an anomaly matching condition. This is made explicit by Witten's theorem, 
 which states that quantum systems described by certain non-trivial TQFTs must exhibit edge modes described by CFTs. It is also known that certain scaling exponents, conformal spins and the central charge modulo 8 of rational CFTs are in one to one correspondence with the quantum dimensions and topological spins of the corresponding anyon theory~\cite{Moore1989,kitaev2006anyons}.  

Non-chiral TQFTs in (2+1) dimensions can be realized on the lattice using the Turaev-Viro state sum construction~\cite{Turaev1992,barrett1996invariants}, or the dual string-net construction~\cite{Levin2005}. The latter provides a physical mechanism which clarifies how topological phases can emerge from microscopic degrees of freedom through the condensation of extended ``string-like'' objects. These models are currently subject to intense study in the fields of condensed matter and quantum information as they describe quantum systems exhibiting topological quantum order, whose physical realizations could potentially yield fault tolerant quantum memories~\cite{qdouble}. The different topological sectors (anyonic excitations) in these systems are given by the Drinfeld center of the input fusion category~\cite{drinfeld,JOYAL199143,majid1991representations,DrinfeldCenter}, which can be found concretely using Ocneanu's tube algebra~\cite{ocneanu1994chirality,tubealgebra,evans1998quantum}.

Rational (1+1)-dimensional CFTs can be realized on the lattice as the scaling limit of critical statistical mechanical models in two dimensions, with the most famous example being the classical critical Ising model. As shown in detail in a series of papers by Fr\"ohlich, Fuchs, Runkel, and Schweigert, the properties and fusion algebra of topological conformal defects in CFTs are very much related to those of topological sectors in $(2+1)D$ TQFTs~\cite{Felder2000,Fuchs2002,Frohlih2004}. This connection has been established directly on the lattice, using string-net models, in a series of papers by Aasen \emph{et al.}~\cite{Aasen2016,daveprep}. 

The aim of this paper is to demonstrate that quantum tensor networks provide a useful 
lens through which one can view 
this correspondence. The main ingredients are as follows: 
first, we make use of the fact that string-nets and their symmetry-enriched (SE) cousins~\cite{PhysRevB.94.235136,cheng2016exactly} have a natural representation in terms of projected entangled pair states (PEPS), and that their topological features are completely characterized by symmetries of the local tensors in the form of matrix product operators (MPOs) \cite{Sahinoglu2014,williamson2014matrix,Bultinck2017,Williamson2017}. The emergent topological sectors are given by the idempotents of a $C^*$ algebra constructed from these symmetry MPOs, which is a representation of Ocneanu's tube algebra~\cite{ocneanu1994chirality,tubealgebra,evans1998quantum}. 
Second, we use a generalized version of the concept of strange correlators (SCs), introduced in \cite{You2014}, where a classical partition function is defined as the overlap between a lattice realization of a symmetry protected topological (SPT) state and a product state with the same global symmetries. We define such strange correlators for long-range entangled string-net wave functions. Naively, one may expect that the overlap between two states of zero correlation length is uninteresting, but the opposite is true: this partition function captures the physics at the interface between a topological phase and a trivial phase, and hence exhibits interesting boundary phenomena. Looking ahead, the topological properties of the string-net ensure that (a subset of) the non-local symmetries, which emerge in the scaling limit of the classical partition function at criticality, are already enforced at the ultraviolet level~\cite{Aasen2016}. 
The third ingredient needed is a folklore structure theorem for nontrivial MPO algebras, which states that any one-dimensional quantum Hamiltonian or transfer matrix on a spin chain which commutes with all elements of such an algebra has to either be gapless/critical or spontaneously break the symmetry. This implies that the partition function obtained using the strange correlator construction (for anything but untwisted gauge theory) will be critical or symmetry broken, thereby reducing the amount of fine-tuning necessary to obtain a critical statistical mechanical model. 

Our tensor network construction is useful for several reasons. 
First, it demonstrates that MPO algebras provide a systematic description of the non-local lattice symmetries that underlie topological properties of emergent CFTs~\cite{Aasen2016}. 
Second, if the input category corresponds to the full Moore-Seiberg data of the emergent CFT and none of the MPO symmetries are broken, it yields a direct way of constructing all different conformal blocks, which is typically a daunting task for critical lattice systems \cite{Mong2014,Aasen2016,Hauru2016}. The strange correlator construction provides a method to obtain those blocks directly from the topological sectors in string-net wavefunctions, and the fusion and braiding of those defects proceeds exactly as in the topological case. We use this approach to numerically identify topological sectors in finite-size CFT spectra of twisted partition functions on the torus. Third, as the strange correlator construction is an overlap between two states with zero correlation length it provides a complementary perspective on real-space renormalization group (RG) schemes, which can now be carried out directly on these quantum states in a fashion that manifestly preserves the MPO symmetries. 

This work has strong relations with the work of Petkova and Zuber \cite{Petkova2000,Petkova2001}, where an algebraic approach was used to construct boundary conditions in CFTs, and with the approach of Aasen, Fendley and Mong~\cite{Aasen2016}, where topologically invariant defects were constructed from defect commutation relations on the lattice. The generalized strange correlator construction is a Euclidean spacetime counterpart to anyonic chain models, where topological symmetries and defects have been discussed previously~\cite{PhysRevLett.98.160409,PhysRevLett.103.070401,Buican2017}. 
The class of partition functions and topological defects produced by the generalized strange correlator matches those considered in Refs.\cite{Aasen2016,daveprep}. 
Viewing these lattice topological defects as MPO algebras leads to a natural numerical implementation via tensor networks and also points towards many possible generalizations such as the case of non-unitary CFTs and higher spatial dimension~\cite{walker2012,Williamson2016,higherdto}.

\simplesection{String-nets, MPO algebras and strange correlators}
From the point of view of tensor networks, string-nets correspond to non-injective PEPS with MPO symmetries on the virtual level: the strings formed by these MPOs satisfy the pulling through equation, and hence are locally invisible on the physical level \cite{PhysRevB.79.085118,PhysRevB.79.085119,Sahinoglu2014,Bultinck2017}. Additionally, these MPOs form a closed algebra that is a representation of the fusion ring $\mathrm{MPO}_{a}\mathrm{MPO}_{b}=\sum_{c}N_{ab}^{c}\mathrm{MPO}_{c}$. The coefficients $N_{ab}^c$ correspond to the fusion data of the input category, while the string-net is completely specified by its $F$-symbols which are solutions to the pentagon equation arising from this input fusion data \cite{Levin2005}. Pairs of anyons can be constructed by putting defect tensors at the end points of MPO strings, the action of the MPO symmetries on these defects defines another $C^*$ algebra whose central idempotents correspond to the topological sectors (anyons) of the  Drinfeld center output category \cite{Bultinck2017,Williamson2017}. The same idempotents yield a basis of minimally entangled ground states for a string-net defined on a manifold with non-trivial genus, and will correspond to the different conformal blocks in the statistical mechanical model if it is critical and no symmetry is broken. 
A SE string-net has a graded algebra of MPO symmetries whose nontrivial components are only free to move up to a physical group action. This leads to a modification of Ocneanu's tube algebra that produces the relative Drinfeld center as a graded output category~\cite{Williamson2017}. 

By projecting the topological PEPS wave function onto a product state $|\omega\rangle^{\otimes N}$, the tensor network representation of a classical partition function is obtained (albeit without a positivity guarantee for the Boltzmann weights); such a construction was first envisioned in the context of strange correlators~\cite{You2014}. A crucial ingredient is the fact that the corresponding row-to-row transfer matrix inherits the MPO-symmetry of the underlying string-net. 
Commuting an MPO in sector $g$ through the transfer matrix leads to an action on the product state $(U_g |\omega\rangle)^{\otimes N}$ which may correspond to a nontrivial duality between phases. When the product state is duality symmetric, the resulting transfer operator will have an enlarged symmetry algebra. 
It was proven in Ref.~\cite{Chen2010} for the MPO representations of a finite group $\mathsf{G}$, which are characterized by the third cohomology group $\mathcal{H}^{(3)}(\mathsf{G},\mathsf{U}(1))$, that no injective MPS can be found in the invariant subspace if the associated 3-cocycle of the MPO representation is non-trivial, i.e.\ if the MPO group action is anomalous~\cite{bridgeman2017anomalies}. This implies that the boundary of a non-trivial SPT phase cannot be gapped with a unique ground state, leaving criticality and symmetry breaking as the only possibilities. This behavior is inherited by the row-to-row transfer matrix of the corresponding strange correlator. 
Analogously, it readily follows from the fundamental theorem of MPS~\cite{Cirac2017100} that an injective MPS cannot be invariant under the action of an MPO algebra that corresponds to a fusion category with nontrivial $F$-symbols (which must be the case if any of the quantum dimensions are greater than one). 
 The non-existence of an injective MPS as unique fixed point of the row-to-row transfer matrix indicates either long range order or power law decay of correlations, exactly as in the SPT case, without any need for fine tuning \footnote{We remark that for string-net models, a symmetry-breaking state corresponds to a unique gapped boundary, since it is projected onto the MPO invariant subspace by the symmetry constraint from the bulk. In the resulting statistical mechanical model, however, purely virtual degrees of freedom in the tensor network are identified with the classical degrees of freedom, such that ``gauge symmetry-violating'' perturbations do become physically accessible.}. We expect that the latter possibility corresponds to the lattice realization of a CFT.
\simplesection{Fibonacci/hard-hexagon model}
\begin{figure}[b]
\centering
\includegraphics[width=0.85\linewidth,keepaspectratio=true]{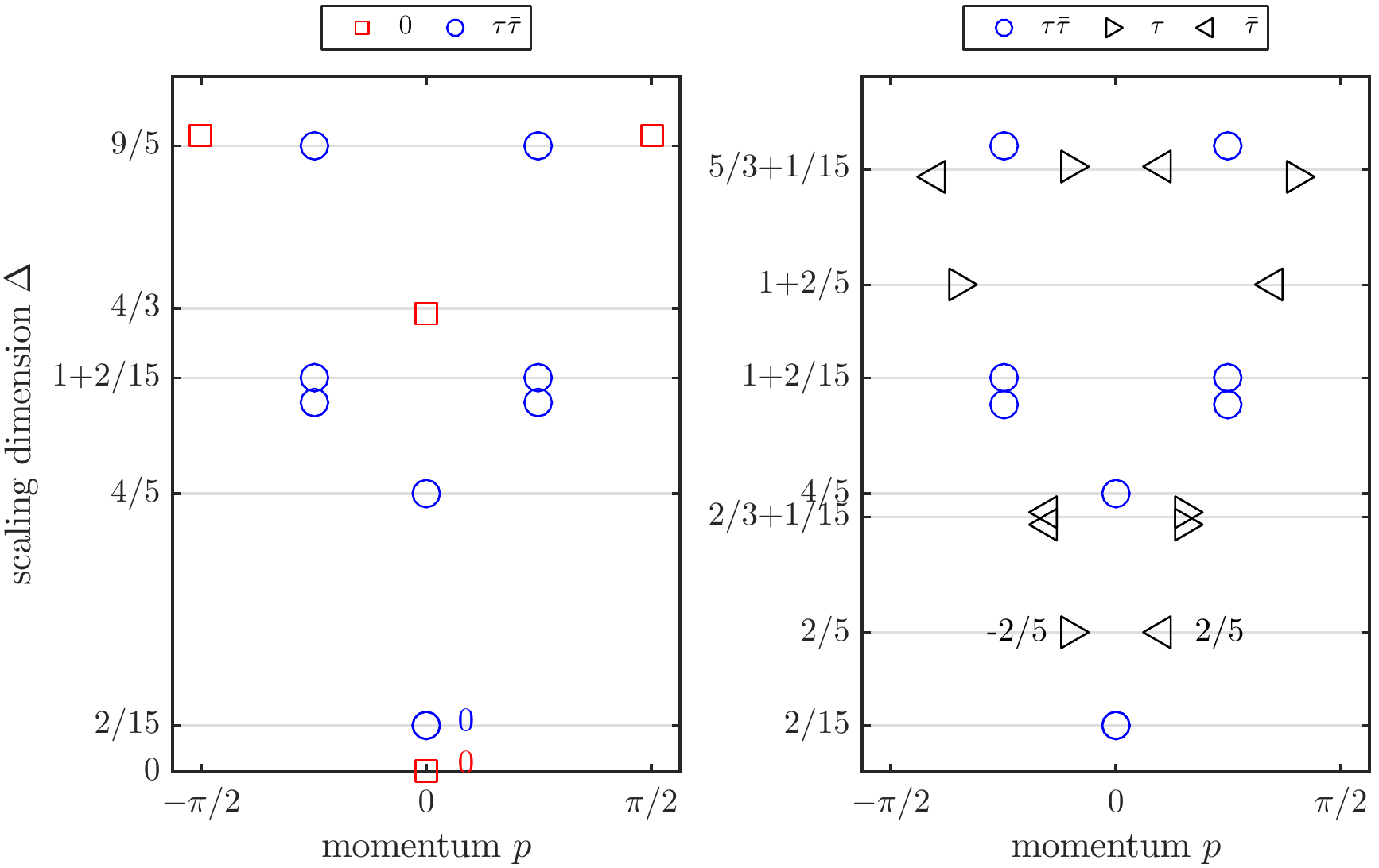} 
\caption{Topological sector labeling of finite-size CFT spectra (scaling dimension $\Delta$ versus momentum $p$) of twisted hard hexagon partition functions on a cylinder (extrapolated from $L_{y} = 18,21,24$). The exact topological correction to the conformal spin is denoted next to the first appearance of the respective idempotents.}
\label{fig:fibonaccispectra}
\end{figure}

As a first example, let us start with the Fibonacci string-net defined on the hexagonal lattice with the objects $\{1,\tau\}$, the non-trivial fusion rule ${\tau \times \tau = 1 + \tau}$ and non-trivial $F$-symbols
\begin{align}
[F^{\tau\tau\tau}_{\tau}]_{ij}=\frac{1}{\phi}\begin{pmatrix}
1 & \sqrt{\phi}\\
\sqrt{\phi} & -1
\end{pmatrix}.
\end{align}

A strange correlator is obtained by projecting all physical degrees of freedom onto the $\tau$-label, which gives rise to the partition function constructed from the tensors 
\begin{align}
\vcenter{\hbox{\includegraphics[width=0.15\linewidth]{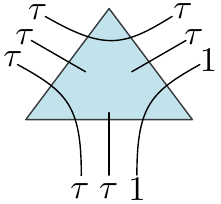}}} = \phi^{1/2} \quad \text{and} \quad
\vcenter{\hbox{\includegraphics[width=0.15\linewidth]{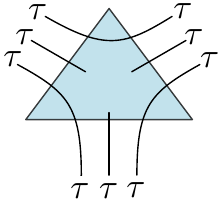}}} = -(\phi)^{-1/2},
\end{align}
and their images under rotation. This tensor network is equivalent to that of the critical hard hexagon model with $c=4/5$ \cite{baxter1980hard,cardy1986operator,fidkowski2009string,fidkowski2007new}, where the internal $\{1,\tau\}$-loops correspond to the presence and absence of a particle in the classical model, respectively. It is quite amazing that this simple ansatz gives rise to the critical fugacity $z_c = \phi^5$. There is a subtlety related to the fact that some Boltzmann weights are negative: by making use of the Euler equation, it can readily be seen that the number of negative terms has to be even, and hence from the point of view of the partition function we can as well ignore the sign. However, the sign has a huge effect on the MPO symmetries in the tensor network: the MPO symmetries are only present for the original F-symbols. 

In Refs.~\cite{Bultinck2017,fidkowski2009string}, the topological sectors of the doubled Fibonacci model were derived in terms of idempotents. Exactly the same idempotents can be inserted into the hard hexagon model, giving rise to the different conformal boundary conditions.  \Fig{fig:fibonaccispectra} shows the exact diagonalization spectra of the transfer matrices $T_{1}$ and $T_{\tau}$ containing respectively no defect line and a $\tau$-defect implemented by the corresponding MPO$_{\tau}$. Due to heavy finite-size effects, the scaling dimensions have been extrapolated using finite-size scaling. The transfer matrices have been projected onto the different topological sectors given by the central idempotents and their eigenvectors have been labeled by the momentum eigenvalues of the twisted three-site~\footnote{The configuration of maximally occupied sites can only be attained for system sizes $L=3n$, $n \in \mathbb{N}$ \cite{Tanaka:2012}.} translation operator \cite{Haegeman2014shadows}. The topological corrections to the conformal spins are enforced exactly by the topological sectors~\cite{Bultinck2017,daveprep,Aasen2016}. The spectra are consistent with the defect partition functions $Z_{1|1}$ (trivial defect) and $Z_{1|9}$ ($\tau$ defect) of the Potts minimal CFT in Ref.~\cite{Petkova2000}.

\begin{figure}[t]
\centering
\includegraphics[width=1.0\linewidth,keepaspectratio=true]{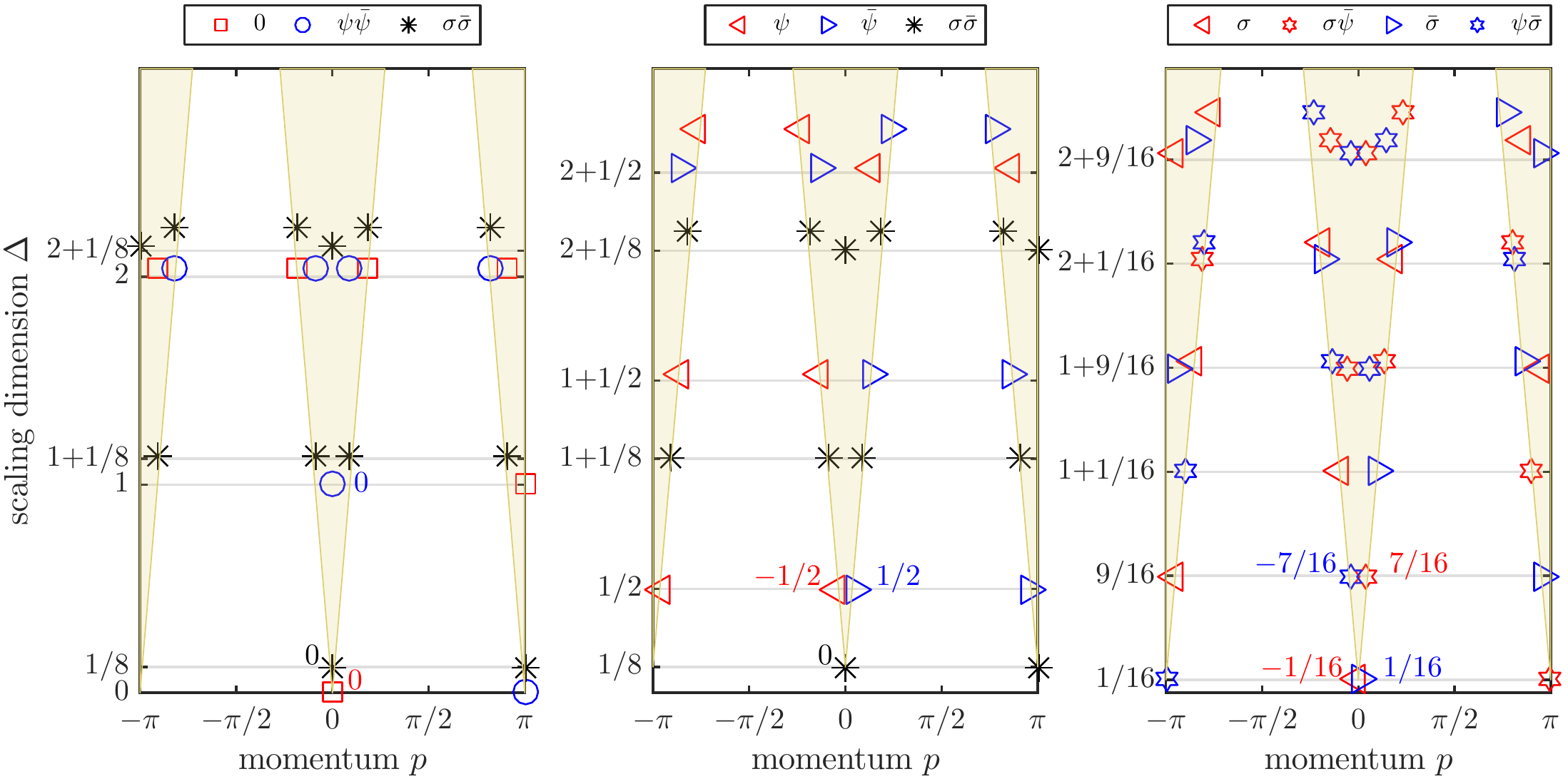} 
\caption{Topological sector labeling of finite-size CFT spectra (scaling dimension $\Delta$ versus momentum $p$) of twisted Ising partition functions on a cylinder ($L_{y} = 11$). From left to right: the $1$-twist, $\psi$-twist, and $\sigma$-twist spectra. The exact topological correction to the conformal spin is denoted next to the first appearance of the respective idempotents.}
\label{fig:isingspectra}
\end{figure}

\simplesection{Ising model} Next, we construct the partition function $Z=\sum_{\braket{ij}} \exp \left(-\beta \sigma_{i}\sigma_{j}\right)$ of the Ising model on the square lattice from the SE string-net wave function based on the Ising fusion category, and insert all 9 possible conformal boundary conditions~\cite{Aasen2016}. This string-net is built from the Ising fusion category with a $\mathbb{Z}_2$-grading on the objects $\{1,\psi\}\oplus\{\sigma\}$ and non-trivial $F$-symbols
\begin{align*}
[F^{\sigma\sigma\sigma}_{\sigma}]_{ij}=\frac{1}{\sqrt{2}}\begin{pmatrix}
	1 & 1\\
	1 & -1
	\end{pmatrix},~ [F^{\sigma\psi\sigma}_{\psi}]_{\sigma}^{\sigma}=[F^{\psi\sigma\psi}_{\sigma}]_{\sigma}^{\sigma}=-1.
\end{align*}
From the PEPS representation of SE string-net wavefunctions~\cite{Williamson2017}, we construct the transfer matrix unit cell obtained from the strange correlator as a direct sum,
\begin{align}
	\vcenter{\hbox{\includegraphics[width=0.24\linewidth]{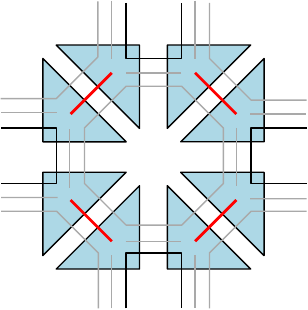}}}\; \bigoplus\;\vcenter{\hbox{\includegraphics[width=0.24\linewidth]{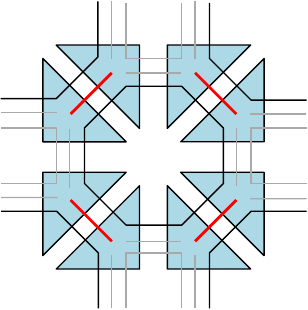}}}\; ,
\end{align}
where the gray lines are fixed to $\sigma$, the black lines to $1/\psi$, and the  red physical indices are acted upon with the product state $\bra{\omega(\beta)}= \sqrt{2}(\cosh(\beta)\bra{1} + \sinh(\beta)\bra{\psi})$. We interpret the virtual $1/\psi$-loops as fluctuating Ising spin variables living on the primal lattice and the fixed $\sigma$-loops as empty dual lattice sites. 
The plaquette variables of the SE string-net allow us to pick out a single copy of the Ising model (projecting onto either the left or right component in the equation above), pulling through the duality $\text{MPO}_\sigma$ flips this choice. 
By enforcing the duality symmetry, and that the partition functions on both primal and dual lattice are equal, MPO$_\sigma$ is lifted to a virtual symmetry which interchanges two shifted copies of the self-dual Ising partition function, yielding the critical temperature $\beta_c=\log(1+\sqrt{2})/2$. This construction result in a transfer matrix with the full Ising category of MPO symmetries which enables us to make use of the central idempotents of the MPO algebra to characterize the different topological sectors on the lattice \cite{Bultinck2017,Williamson2017}. If we did not promote $\text{MPO}_\sigma$ to a symmetry by considering both primal and dual lattices we would achieve a weaker sector decomposition~\cite{Aasen2016,Hauru2016} where $\text{MPO}_\sigma$ is treated as an extrinsic defect~\cite{Williamson2017}. 
Inserting a $\psi$-twist along a cylinder induces anti-periodic boundary conditions by effectively flipping spins across the defect line. Inserting a $\sigma$-twist implements twisted boundary conditions corresponding to the Kramers-Wannier duality. 

In \Fig{fig:isingspectra}, we plot the twisted finite-size CFT spectra for the Ising model. We remark that the $\sigma\bar{\sigma}$ sector appears in both $T_{1}$ and $T_{\psi}$ since its central idempotent contains contributions from basis states involving both twists, which are related by the $\text{MPO}_\sigma$ symmetry. For the branches around momentum $p=0$, the topological corrections to the conformal spins \footnote{Since a twisted half-shift translation operator was used, the conformal spins $s_{a,i}$ are respectively related to the momenta $s_{a,i}$ by $s_{1,i}=2L_{y}p_{1,i}/(2\pi)$, $s_{\psi,i}=2L_{y}p_{\psi,i}/(2\pi)$, and $s_{\sigma,i}=2(L_{y}-1/2)p_{\sigma,i}/(2\pi)$.} are consistent with the topological spins of the anyons. Due to the superposition of the primal and dual lattice, the eigenvalues of the transfer matrix have an additional degeneracy, which is the origin of the spurious fields appearing around momentum $p=\pm\pi$. The labeling of these spurious fields is compatible with the presence of an additional $\psi\bar{\psi}$ vacuum at momentum $p=\pm\pi$ corresponding to an antisymmetric combination of states from the two lattices. Indeed, $\psi\bar{\psi}$ appears with scaling dimension $\Delta=0$ in the untwisted partition function spectrum \footnote{To fully promote $\text{MPO}_\sigma$ to a symmetry, introducing both lattices is inevitable since the sector idempotents contain tubes with $\mathrm{MPO}_{\sigma}$, mixing primal and dual lattices, i.e.~some emergent anyons are inextricably made up of primal and dual objects, or, in terms of the classical Ising interpretation, of spin operators and disorder operators \cite{Kadanoff1971}.}.

A strange correlator representation of the classical partition function of the 3-state Potts model can be obtained by using the $\mathbb{Z}_{3}$ Tambara-Yamagami category~\cite{TAMBARA1998692}. Taking this data as the input category and constructing an appropriate product state $\ket{\Omega}$ we immediately find a representation of the 3-state Potts model partition function. The critical SC produced in this way contains only a subset \footnote{The CFT describing the 3-state Potts model is actually not the bare $c=4/5$ minimal model, but a related model whose modular invariant partition function involves only a subset of the $c=4/5$ primary fields with different multiplicities. These multiplicities indicates that the 3-state Potts model is not just a subtheory of the $c=4/5$ minimal model, as it contains copies of some of its fields. Similarly, the 3-state Potts fusion rules are not just a subset of those of the $c=4/5$ minimal model \cite{DiFrancesco1997}.} of the primaries of the unitary minimal model with central charge $c=4/5$.

More generally, finding the input unitary fusion category that leads to the full set of topological sectors requires a priori knowledge of the emergent CFT.  If one uses a category that is Morita equivalent to the Moore Seiberg category derived from a chiral half of the emergent CFT, then through the strange correlator construction, the partition function representation on the lattice contains the same topological data as the emergent CFT. For the lattice sectors to truly match those of the CFT, we must further require that none of the MPO symmetries is spontaneously broken. In that case properties of the tube algebra ensure that the topological spins and primary scaling dimensions are built-in and have to exactly match those of the resulting CFT~\cite{Aasen2016,daveprep}. 

\simplesection{Real-space renormalization}
Having discussed the utility of the strange correlator perspective for characterizing the topological sectors appearing in CFT spectra, let us now discuss its real-space renormalization properties. Since we have decomposed the classical partition function into the overlap of two quantum wave functions (one of them a product state), we can reinterpret known tensor network renormalization procedures on the partition function \cite{Levin2007,Evenbly2015,Yang2017,Bal2017,Hauru2017} at the level of these quantum states. This approach is particularly appealing since string-nets are known to be exact zero correlation length RG fixed-point wave functions under an isometric quantum circuit whose gates are constructed from $F$-symbols~\cite{Konig2009,koenig2010quantum,higherdto}. Each step of this coarse-graining circuit is a projected-entangled pair operator (PEPO) $U$, and inserting $U$ and its conjugate between the SC overlap $\braket{\Omega | \Psi_{\mathrm{SN}} }$ preserves its value. This immediately results in a sequence of effective partition functions induced by the exact renormalizability of the string-net since ${\braket{\Omega^{(i)} | U^\dagger U | \Psi_{\mathrm{SN}}^{(i)} } = \braket{\Omega^{(i)} | U^\dagger | \Psi_{\mathrm{SN}}^{(i+1)} } \simeq \braket{\Omega^{(i+1)} | \Psi_{\mathrm{SN}}^{(1)} }}$, where $\bra{\Omega^{(i+1)}}$ is a truncation of the PEPS $\bra{\Omega^{(i)} } U^\dagger$, and $\ket{ \Psi_{\mathrm{SN}}^{(i+1)} }$ is a string-net on a coarse grained lattice. 
While the initial application of $U$ entangles the product state $\bra{\Omega}$ to give a PEPS that can be stored exactly, further iterations require truncation if we want to cap the virtual bond dimension at a fixed value. By acting with the string-net RG circuit and its conjugate inside the SC $\braket{\Omega | \Psi_{\mathrm{SN}} }$, we are thus able to gradually shift the renormalization group scale. We remark that this introduces correlations into the the input state $\bra{\Omega}$, which initially pertained to the ultraviolet physics of the critical lattice model, i.e.~the Boltzmann weights which tune the lattice model to criticality, but leaves invariant the universal part $\ket{\Psi_{\mathrm{SN}}}$, which contains the topological data of the RG fixed point model. This ensures that the MPO symmetry is preserved exactly throughout the renormalization flow. 

It would be very interesting to understand the highly entangled states $\bra{\Omega^{(\infty)}}$ that are exact fixed points under the renormalization circuit and yield nontrivial CFT partition functions, the same data would specify gapless boundary conditions for Turaev-Viro TQFTs. In the case of gapped fixed points, and boundary conditions, $\bra{\Omega^{(\infty)}}$ is specified by an algebra object in the input fusion category. 

\begin{figure}[t]
\centering
\includegraphics[width=0.65\linewidth,keepaspectratio=true]{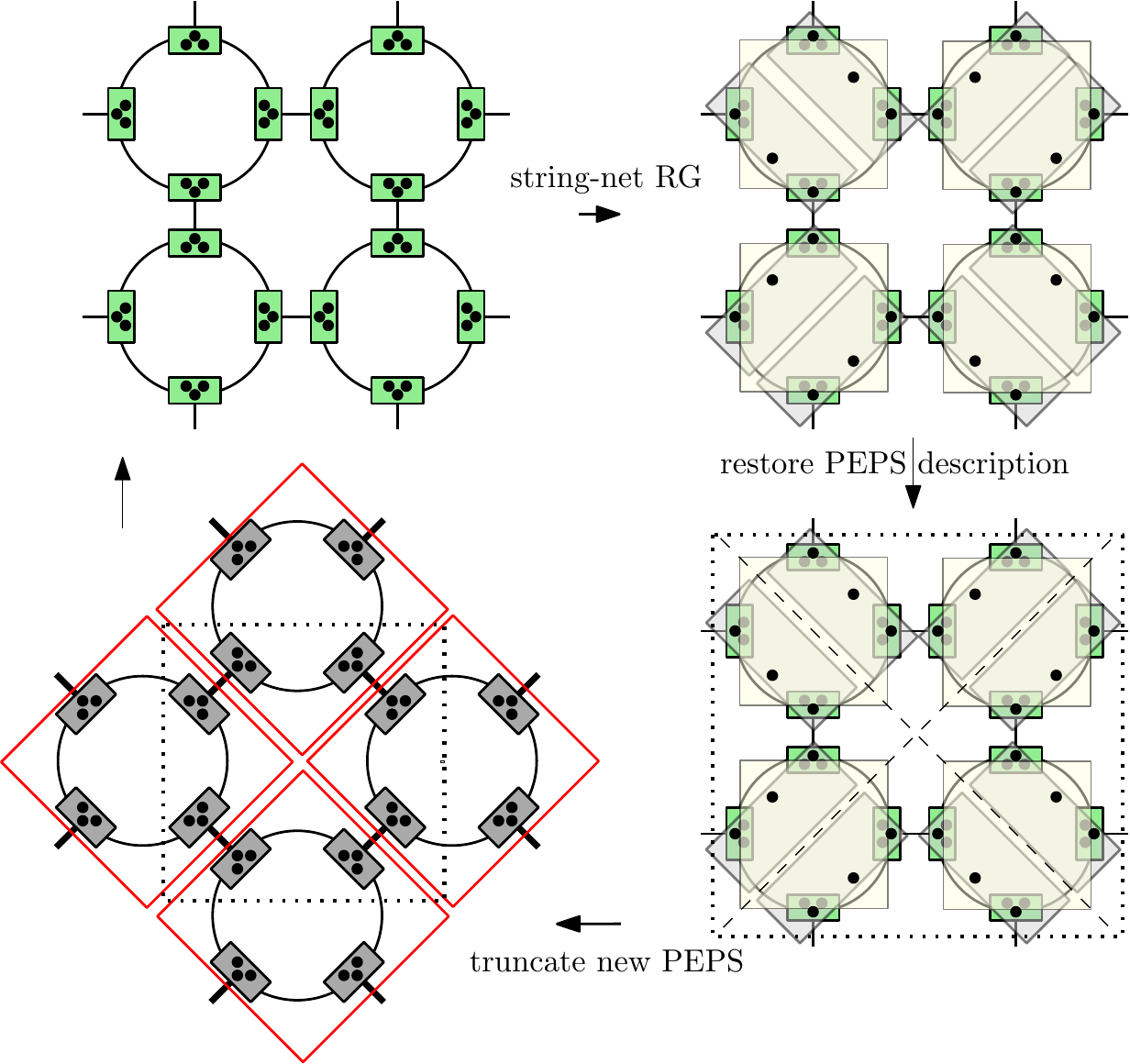} 
\caption{Every RG step, the gates of the exact string-net coarse-graining map block the PEPS tensors in such a way that preserving the PEPS structure leads to an increased bond dimension, indicating that the coarse-grained PEPS has to be truncated in order to obtain a sustainable RG transformation.}
\label{stringnetrg:fig:peps_trg_tnr}
\end{figure}

In \Fig{stringnetrg:fig:peps_trg_tnr}, we depict the internal substructure of the coarse-grained PEPS at any layer in terms of four tensors which can be derived from those of the previous layer. Some of the gates in $U$ can be done on each of these four tensors individually, but other gates connect the coarse-grained PEPS with its nearest-neighbor tensors and require truncation. Crucially, it is the way in which this truncation step is carried out which differentiates between different tensor network renormalization schemes, since the action of the coarse-graining PEPO on the physical level of the PEPS input state amounts to (i) blocking sites, which increases the virtual bond dimension, (ii) projecting out half of the physical indices, and (iii) constraining the surviving physical indices of the PEPS to configurations allowed by the string-net fusion rules. The actual renormalization step, which gets rid of local, short-range correlations, depends solely on the choice of PEPS truncation on the virtual level. This leads to effective partition function tensors at every layer given in terms of truncated, renormalized strange correlators
\begin{align}
	\vcenter{\hbox{\includegraphics[width=0.45\linewidth]{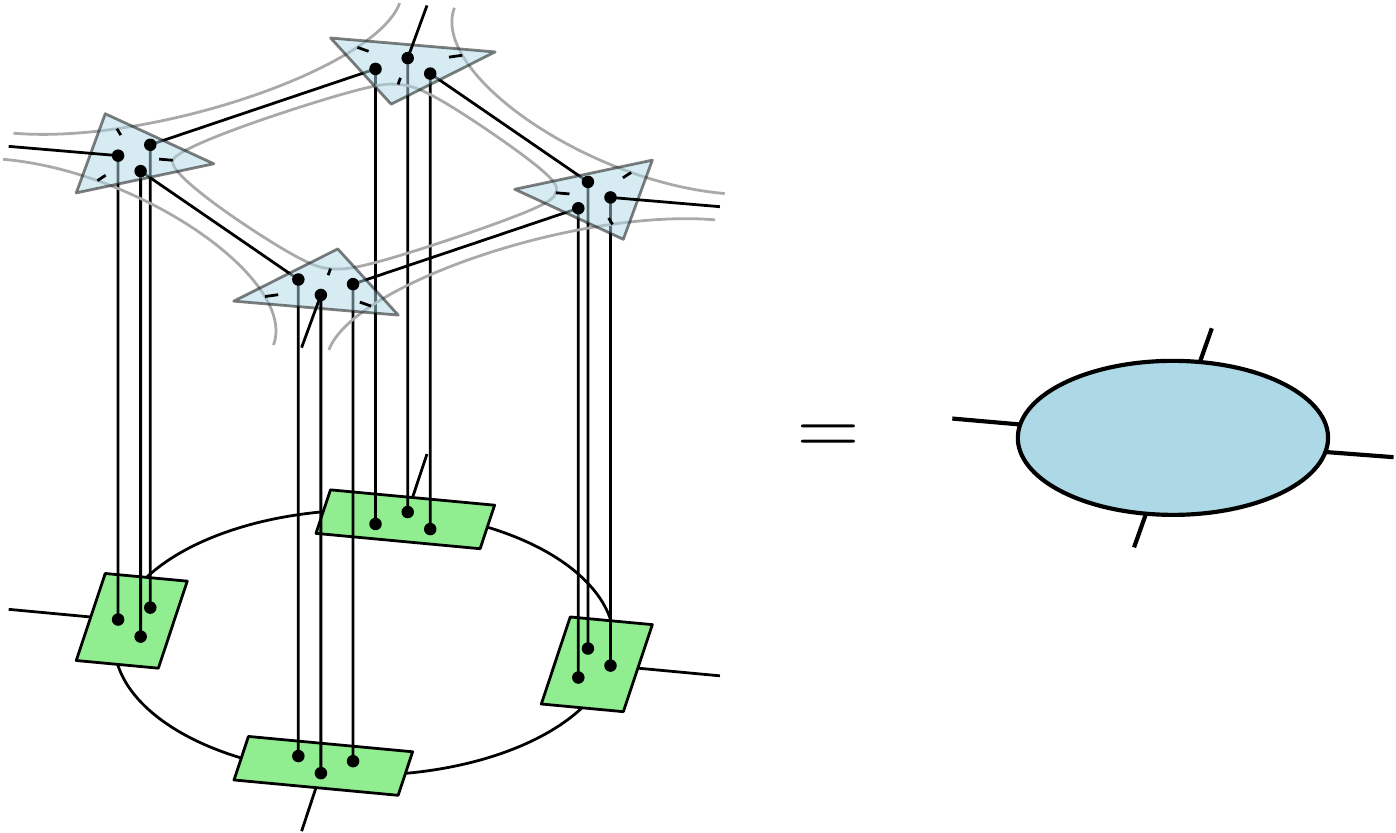}}}\label{eq:pepsscoverlap}
	\, .
\end{align}
By enforcing that the input state PEPS remains a product state, we obtain a flow that remains within the Ising phase diagram, reminiscent of Kadanoff's block-spin method truncated to nearest-neighbor interactions. The simplest non-trivial way to truncate the input PEPS is by doing independent singular value decompositions to pull the coarse-grained tensors apart in \Fig{stringnetrg:fig:peps_trg_tnr}. We have observed numerically that this SC-TRG truncation procedure leads to a flow of partition functions whose free energy accuracy and fixed point behavior is indistinguishable from that of the tensor renormalization group (TRG) algorithm \cite{Levin2007}. In terms of PEPS optimization, the most straightforward improvement is to include the full environment and set up a variational PEPS optimization which maximizes the overlap between the coarse-grained PEPS with an increased bond dimension and a truncated one. However, we have observed numerically that the full PEPS environment is not optimal for the removal of local, short-range correlations in $\bra{\Omega^{(i)}}$ as part of a coarse-graining algorithm. This is consistent with the success of TNR methods, all of which remove local correlations in a highly asymmetric way \cite{Evenbly2015,Yang2017,Bal2017,Hauru2017}. We leave the design of such a PEPS truncation procedure for future work but already note the relevance of the recently developed implicitly disentangled renormalization \cite{Evenbly2017}.

\simplesection{Conclusions} We have generalized the concept of strange correlators to construct critical partition functions on the lattice starting from string-net wavefunctions. The defining feature of the string-nets are the matrix product operator symmetries characterizing their nonlocal entanglement structure, 
and we have used those same MPOs to construct the topological sectors of the emergent CFTs. 
We also exploited the real space renormalization group fixed point nature of the string-nets to define a symmetry-preserving coarse-graining procedure on the strange correlator partition functions. 
We expect that all rational CFTs can be obtained through the strange correlator construction, and that all of their conformal blocks can be found by means of the corresponding tube algebras (subtle issues that may arise are dealt with more comprehensively in the series of papers Refs.~\cite{Aasen2016,daveprep}). In particular, it is known that all minimal models can be realized in the closely related anyon chain models~\cite{PhysRevLett.98.160409,PhysRevLett.103.070401}. The tensor network and matrix product operator approach is not inherently limited to 2-dimensional systems or unitary fusion categories, and hence it is an intriguing prospect to apply it to higher dimensional and non-unitary CFTs~\cite{inpreparation}. It would also be interesting to apply the strange correlator method to fermionic tensor networks and dualities~\cite{bultinck2016fermionic,williamson2016fermionic,NewFermionPEPSPaper2017,PhysRevB.95.245127,aasen2017fermion}.

\simplesection{Acknowledgements}
We would like to thank D. Aasen, J.~Bridgeman, G.~Evenbly, P. Fendley and L.~Vanderstraeten for inspiring discussions. DW especially thanks D. Aasen for explaining his work~\cite{Aasen2016,daveprep}, particularly the relevance of Ocneanu's tube algebra to topological sectors in CFTs. This work is supported by an Odysseus grant from the FWO, ERC grants QUTE and ERQUAF, and ViCoM and FoQuS.

\bibliographystyle{apsrev4-1}
\bibliography{strangecorrelator}


\end{document}